\documentclass[journal]{IEEEtran}
\IEEEoverridecommandlockouts
\usepackage{cite}
\usepackage{amsmath,amssymb,amsfonts}
\usepackage{algorithm}
\usepackage{algorithmic}
\usepackage{graphicx}
\usepackage{caption}
\usepackage{textcomp}
\usepackage{xcolor}
\usepackage{url}
\usepackage{booktabs}
\usepackage{siunitx}

\def\BibTeX{{\rm B\kern-.05em{\sc i\kern-.025em b}\kern-.08em
    T\kern-.1667em\lower.7ex\hbox{E}\kern-.125emX}}
\begin{document}

\title{Federated Learning in Mobile Edge Computing: An Edge-Learning Perspective for Beyond 5G\\
}
\author{\IEEEauthorblockN{Shashank Jere, Qiang Fan, Bodong Shang, Lianjun Li, and Lingjia Liu}
\thanks{The authors are with the Bradley Department of Electrical and Computer Engineering, Virginia Tech, Blacksburg, VA, USA.}
\thanks{The corresponding author is L. Liu (ljliu@ieee.org).}
}

\maketitle

\begin{abstract}
Owing to the large volume of sensed data from the enormous number of IoT devices in operation today, centralized machine learning algorithms operating on such data incur an unbearable training time, and thus cannot satisfy the requirements of delay-sensitive inference applications. By provisioning computing resources at the network edge, Mobile Edge Computing (MEC) has become a promising technology capable of collaborating with distributed IoT devices to facilitate federated learning, and thus realize real-time training. However, considering the large volume of sensed data and the limited resources of both edge servers and IoT devices, it is challenging to ensure the training efficiency and accuracy of delay-sensitive training tasks. Thus, in this paper, we design a novel edge computing-assisted federated learning framework, in which the communication constraints between IoT devices and edge servers and the effect of various IoT devices on the training accuracy are taken into account. On one hand, we employ machine learning methods to dynamically configure the communication resources in real-time to accelerate the interactions between IoT devices and edge servers, thus improving the training efficiency of federated learning. On the other hand, as various IoT devices have different training datasets which have varying influence on the accuracy of the global model derived at the edge server, an IoT device selection scheme is designed to improve the training accuracy under the resource constraints at edge servers. Extensive simulations have been conducted to demonstrate the performance of the introduced edge computing-assisted federated learning framework.

\end{abstract}


\section{Introduction}
The number of IoT devices is expected to increase exponentially in the next decade~\cite{Cisco2017global}. Due to the large number of connected Internet of Things (IoT) devices, the worldwide IP traffic is projected to grow to $400$ Exabytes per month by $2022$.
As the computational and communication capabilities of such devices continue to improve, there is growing interest in harnessing their capabilities for machine learning applications~\cite{Hao2020AI}.
The complex and heterogeneous nature of problems likely to be encountered in the context of next-generation Beyond-$5$G cellular networks, especially with the adoption of modern techniques such as ultra-massive MIMO, millimeter-wave (mmWave) usage, beamforming, etc. is demonstrating traditional model-based techniques to be insufficient in finding optimal solutions to such problems~\cite{RubayetAI}.
Additionally, domain-specific knowledge, which is a characteristic of datasets generated during network operation, can be used to provide for better interpretation of the training data, leading to better trained models~\cite{Zhou2020:inpress-b}.
Based on the data streams generated by IoT devices, some IoT applications such as autonomous vehicles, environment monitoring and augmented reality can employ machine learning methods (such as deep neural networks) to train a learning model to make the right decision timely, thus satisfying their real-time control requirements. Specifically, the remote cloud can collect the data streams generated from various IoT devices and train a global model in a centralized way. Subsequently, the global model will be feedbacked from the cloud to IoT devices for inference tasks. However, owing to the huge volume of IoT devices and the long communications delay, the training time is almost always infeasible for delay-sensitive applications, whose models may be time-varying in the highly dynamic environment. To solve this problem, mobile edge computing, a computational paradigm that places computing resources at the network edge~\cite{MaoMEC2017} is integrated  with the local computing resources of IoT devices to execute training tasks in a decentralized way (i.e., federated learning). Due to the rapid development in hardware platforms for end user devices, each IoT device can employ its local computing resources to train a local model based on its limited local dataset, the parameters of which are then transferred to the edge servers for further processing. After aggregating all model parameters from various IoT devices, the edge server can summarize the model parameters and derive the corresponding global model. Since both the local training tasks and the parameters transfer have a comparatively lower delay, the training time for the global model is significantly reduced.

In Federated Learning (FL), both training time and global model accuracy are crucial metrics. On one hand, the training time for the global model is impacted by the local training process and the communications delay between IoT devices and the edge server. Thus, reducing this communications delay becomes a critical issue. As the wireless channel environments are highly dynamic and local model parameters at different IoT devices are transmitted to the edge server at different times, the available radio resources in the network are time-varying. Meanwhile, as the future traffic load is unpredictable, the radio resource allocations for different traffic tasks (i.e., model parameters) are coupled with each other. In other words, if the system allocates too many wireless channels for one traffic task owing to its desirable channel condition, the subsequent tasks may not get enough radio resources and thus violate the QoS requirement. On the other hand, this may degrade the model accuracy in federated learning since the global model is derived based on local model parameters, contrary to centralized machine learning in the remote cloud. Therefore, global model accuracy remains an important problem in federated learning especially when the computing and storage resources of edge servers are limited.        

To solve the above problem, we design a novel edge computing assisted federated learning framework that aims to improve both the training time and training accuracy in the network. The major contributions are summarized as follows: 
\begin{itemize}
\item We introduce an edge computing assisted federated learning framework and analyze its potential in improving the training efficiency.
\item We present the problem of improving the communications delay incurred in transferring model parameters to edge servers and propose a reinforcement learning based algorithm to solve it in real time.
\item We present the problem of selecting IoT devices for federated learning based on their different weights in enhancing the training accuracy while the computing and storage resources of the edge server are limited.
\end{itemize}

\section{State-of-the-Art}
\label{state_of_art}

\subsection{Multi-Access Edge Computing}
Although mobile cloud computing~\cite{FernandoMCC2013} has been a successful paradigm in bringing richer and more complex applications to end users by harnessing the power of the server cloud, extremely sensitive latency requirements have demanded an alternate approach~\cite{MaoMEC2017}.
Driven by the complex traffic distributions in wireless communications systems, the network architecture is becoming increasingly heterogeneous.
There are multiple types of network access nodes providing reliable and seamless connectivity for mobile users such as macro base station (BS), small cell BS and WiFi access point, to name a few.
These network access nodes support edge computing at network edges with low transmission latency.
Due to the different characteristics of network access nodes such as coverage ability and transmit power, the design of the coexistence of heterogeneous MEC networks has attracted increasing attention~\cite{MaoMEC2017}.
The cooperative computation offloading between multiple different network access nodes for mobile users needs to be well-designed.

Under such a heterogeneous network architecture, the task allocation and resource allocation among different network nodes can significantly improve system performance. On one hand, cooperation between edge computing and cloud computing can be achieved to enhance the quality of service of IoT tasks further. Specifically, cloud servers can process tasks that require intensive computation, while edge servers can process computation tasks that entail a small data size or have a low latency requirement~\cite{fan2018workload}. On the other hand, the task allocation among edge servers can effectively offload tasks from overloaded edge servers to underutilized ones, and thus reduce the execution delay of tasks~\cite{fan2018towards}.



\subsection{A Primer on Federated Learning (FL)}

A traditional cloud-only based machine learning approach offloads data sensed at end devices to the remote cloud server for centralized training in order to learn the model for future inference. However, the training time at the cloud may be impractical owing to the large volume of the sensed data that needs to be utilized in the training process. Meanwhile, as the cloud server may be physically distant from the end devices, these devices may suffer from large communication delays. To solve this problem, federated learning facilitated by MEC can be a promising approach to shift from a centralized training paradigm to a more practical decentralized training one. Federated Learning (FL)~\cite{McmahanFedML2017} aggregates the results of distributed training models at different end devices based on their local datasets, and cooperatively derives the global training model. Specifically, each device can train its local model based on its own limited dataset efficiently and update the local model parameters to an aggregator at the edge server. The aggregator then utilizes each learner's local model parameter updates to optimize a global model. This can be repeated for a number of iterations to ensure the global model reaches a desired accuracy. Subsequently, the global model parameters are fed back to the distributed devices for future inference. 

Federated Learning distinguishes itself from other distributed learning schemes owing to certain unique factors. Firstly, the assumption that the data samples sensed at the end devices of different users are realizations of independent and identically distributed (IID) random variables may not hold in FL, since the local dataset of a single user's end device may not be representative of the overall population distribution. Secondly, the local datasets generated across federated learners may differ greatly in size, causing an imbalanced distribution. 
This imbalance in dataset sizes are primarily due to the different types of IoT devices (e.g., smartphone or vehicle) and different application scenarios (e.g., a maps application on a smartphone may generate more data for an active city user than a less active rural user). 
Thirdly, in the FL setting, the total amount of sensed data samples contributing towards learning the global model at the edge server is much larger than that available for local training at each user. 
Finally, most federated learners are mobile devices (e.g., smartphones, wearables, drones, vehicles, etc.) with possibly unreliable wireless connectivity to the FL edge server. This implies that the aggregator may have to support learners that may be offline or experience slow connectivity. In the context of these aforementioned differentiating factors~\cite{McmahanFedML2017}, FL provides advantages not available in other decentralized ML approaches, namely, user-data privacy, efficient network resource usage and reduced end-to-end latency.

\section{Federated Learning Facilitated by Mobile Edge Computing}
\begin{figure*}[!h]
\centering
    \includegraphics[width=6.0in, height=3.0in]{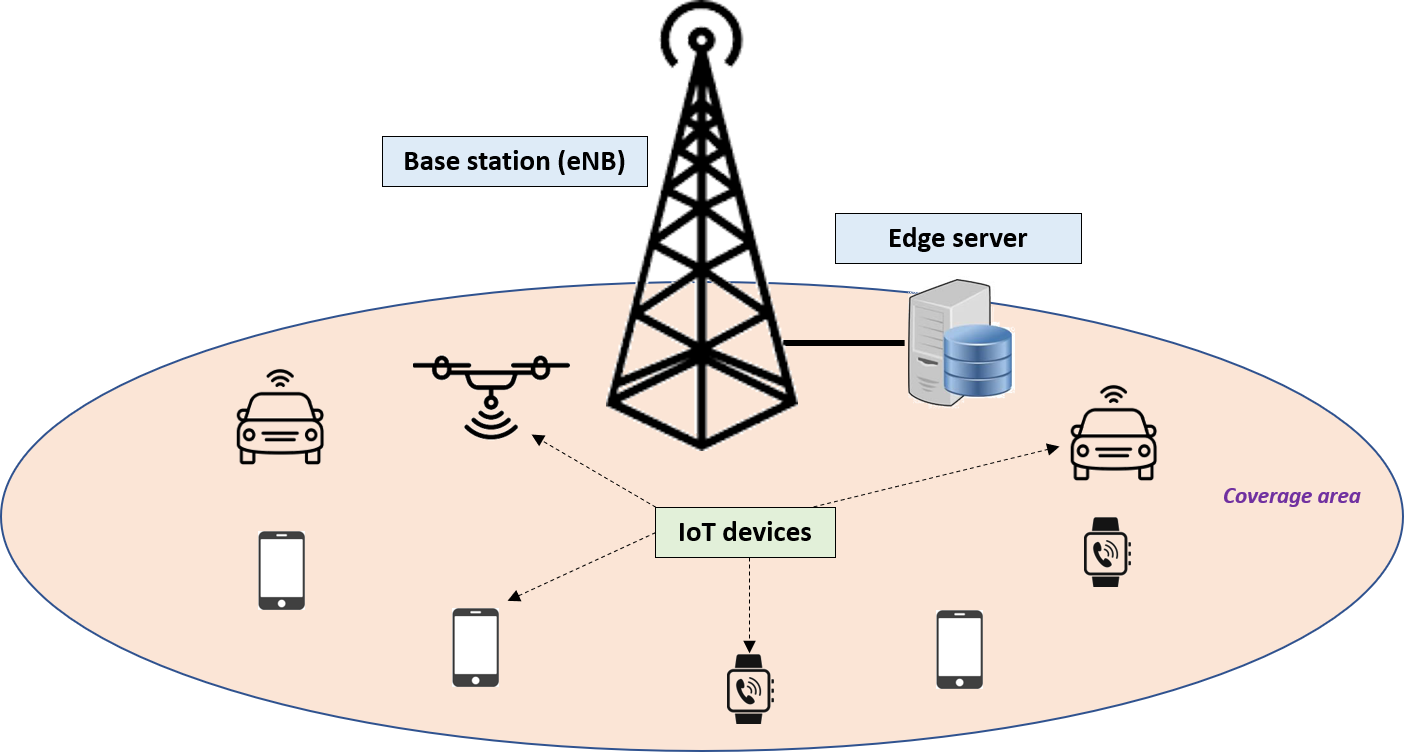}
    \caption{General MEC Architecture}
    \label{general_MEC_fig}
\end{figure*}

FL and MEC have a promising synergy that is particularly well-suited for use cases proposed for Beyond-$5$G. Figure~\ref{general_MEC_fig} describes the basic concept of MEC with a single edge server that provisions services for a large number of IoT devices. 
The Quality of Service (QoS) of the notable delay-sensitive applications such as AR/VR, drone-based communications and Vehicle-to-Infrastructure (V2I) communications, will be heavily impacted by the overall round-trip latency between the edge server and distributed IoT devices. 

In this section, we will present the edge computing-assisted federated learning framework in which both the training efficiency and training accuracy will be taken into account. 
Firstly, minimizing communication latency will accelerate the mutual interactions between IoT devices and the edge server, thus reducing the training time of a global model in various applications. To reduce the communications delay of local model parameters, we investigate this problem in Section~\ref{comm_for_FL}. Secondly, as the network involves heterogeneous devices, the different datasets enable various IoT devices' local model parameters to have different weight on the global model accuracy. Meanwhile, considering the huge number of IoT devices and the limiting processing capability of each edge server, it is challenging to select the suitable IoT devices to optimize the global model accuracy.  This optimization problem that attempts to maximize training accuracy while being constrained by both channel conditions of end devices as well as the computational resources at the edge server is set up in Section~\ref{MEC_assisted_FL}.

\subsection{Communications for Federated Learning}
\label{comm_for_FL}

Under the FL setting, different end devices need to transfer the updated model parameters to an edge server that can finally summarize the global model. In this case, the communications latency becomes an important factor determining the performance of the aggregated model. 
In other words, a low communications latency can enable end devices to transfer their local model parameter updates quickly, and thus reduce the total delay for deriving the global model.

These devices may have different channel conditions towards a BS and thus require different volume of spectrum resources to transfer their traffic tasks (which are generated by local training). Given a traffic task of each end device, we need to focus on two factors: 1) how to allocate spectrum resources to improve the communications latency of all tasks; 2) how to guarantee QoS requirement of different tasks in terms of the maximum communications latency requirement. Since latency is a critical metric for FL, it is desirable to minimize the communications latency for as many traffic tasks as possible. On the other hand, although some IoT devices may have unsatisfactory channel conditions, we still need to allocate the required resources to their tasks to meet their QoS requirement, thus ensuring the reliability of FL in the system. Here, denote $\beta$ as a spectrum resource block, $\eta_{i}$ as the spectral efficiency (bits/sec/Hz) based on the corresponding end device’s channel condition, $x_i$ as the number of resource blocks allocated to task $i$, and $l_{i}$ as the length of task $i$. The communications delay of task $i$ is hence derived as $\frac{{{l_{i}}}}{{x_{i}\beta {\eta_{i}}}}$. Meanwhile, $\tau$ is the maximum delay requirement for a task and $W_0$ is the total bandwidth of the wireless channel. Therefore, the resource allocation problem 
can be formulated as follows: 

	\emph{Given:}
	\begin{itemize}
		\setlength\itemsep{-.2em}
		\item	The length of all traffic tasks.
		\item	The channel conditions of end devices where traffic tasks are generated.
		\item	The transmission power of end devices and the total resource blocks of the system (i.e., the total bandwidth).
		\item    The QoS requirement of different tasks.
	    \end{itemize}
	
	\emph{Obtain:}
	\begin{itemize}
		\setlength\itemsep{-.2em}
		\item	The number of resource blocks allocated to each task.
	\end{itemize}
	
	\emph{Objective:}
	\begin{itemize}
		\setlength\itemsep{-.2em}
		\item Minimize the average communications delay of traffic tasks, while guaranteeing the communications delay requirement of each task.
	\end{itemize}

\begin{figure*}[h!]
\centering
    \includegraphics[width=5.0in, height=3.33in]{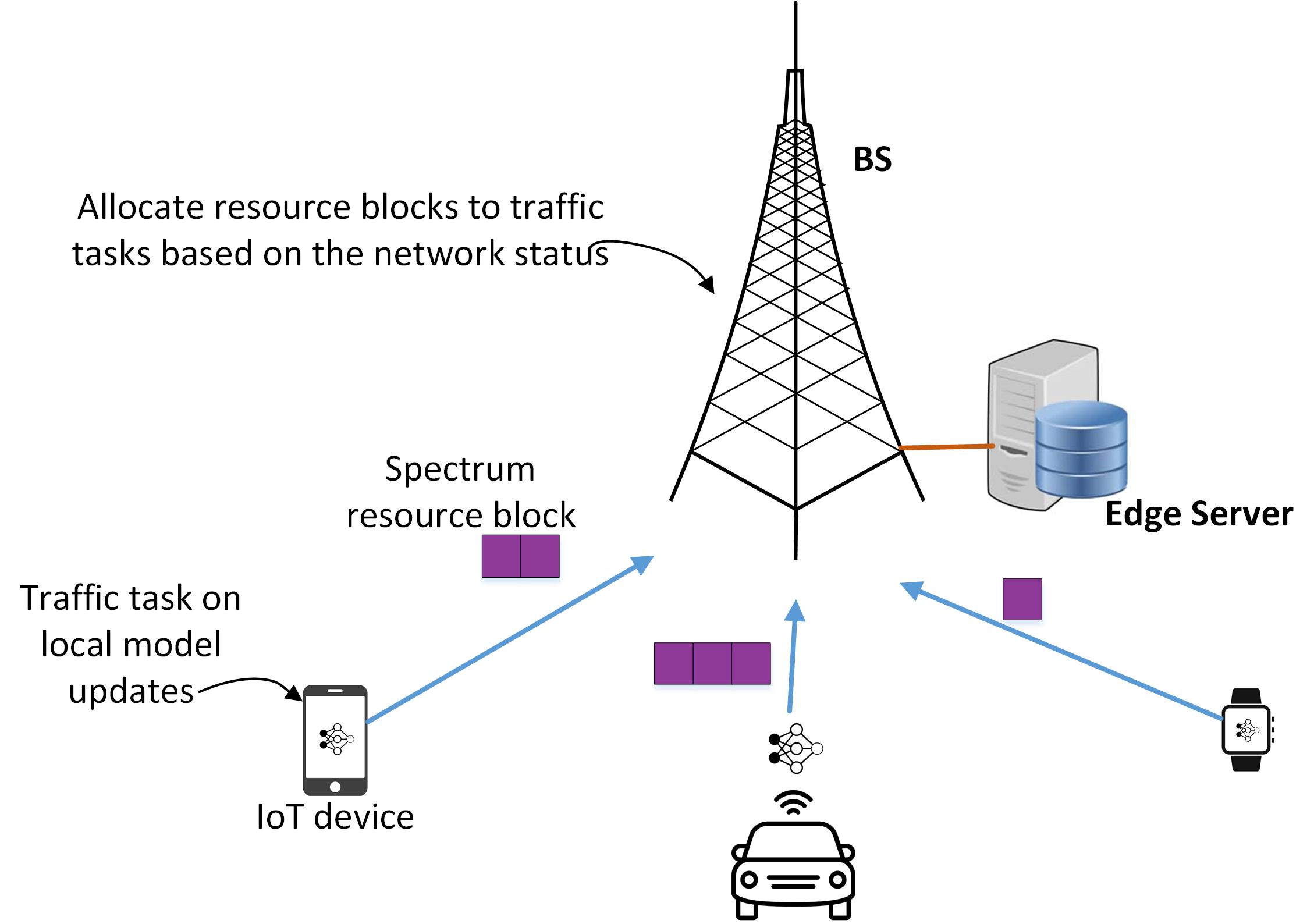}
    \caption{Resource allocation in communications for Federated Learning}
    \label{comm_for_FL_fig}
\end{figure*}

As a mobile end device may visit several locations, it may have time-varying channel conditions and may generate different traffic tasks at different times. Note that the resource allocation for the current task and future tasks are coupled with each other. The optimal resource allocation decision relies on the complete global network information such as task information. 
However, owing to the difficulty of predicting future task information in advance, it is impractical to optimize the resource allocation. Furthermore, as the above resource allocation problem is an integer non-linear problem and the number of end devices may be exceedingly large, it is also challenging to arrive at an optimal solution in real-time. Thus, a data-driven approach such as reinforcement learning (RL) can be adopted to iteratively learn the features of the problem based on historical data and achieve a sub-optimal solution. Specifically, we can design the state of the system to represent 1) the channel condition of IoT devices 2) the available spectrum resource, and 3) the traffic length of the arriving task, while the action can represent the number of resource blocks estimated to be allocated to the arriving task. The agent at the edge server can be trained to learn the policy (i.e., how to conduct the desirable action based on the current state) to maximize the reward (i.e., reciprocal of the average communications delay of tasks). Thus, the system can react to the changing network status quickly and allocate proper resources to different traffic tasks in real-time, thereby improving their communications delay. Figure~\ref{comm_for_FL_fig} depicts the spectrum resource allocation tasks undertaken by the edge server.
 
\subsection{Mobile Edge Computing-assisted Federated Learning}
\label{MEC_assisted_FL}
In this section, the model accuracy of the framework is investigated under the constraints of wireless channels and limited storage and computing resources of edge servers. Since the participating learners include IoT devices such as sensors, drones, autonomous vehicles and smart wearable devices along with smartphones, their ability to upload model state information updates to their corresponding edge servers is highly dependent on their respective channel conditions. Meanwhile, considering the limited storage and computing resources of edge servers, the total number of IoT devices in the network is expected to be much larger than the number of devices that the edge server can accommodate in a given training round. The role of IoT devices in contributing to the global model of FL is vastly different from their roles in generic distributed ML approaches. This is due to their unique distinguishing features which include their unreliable wireless channel conditions, widely-varying local dataset sizes, limited on-device processing capabilities and device energy constraints. The focus here is to introduce an optimal learner selection method in each training round before the global aggregation phase occurs.

\begin{figure*}[!h]
\centering
    \includegraphics[width=6.0in, height=3.0in]{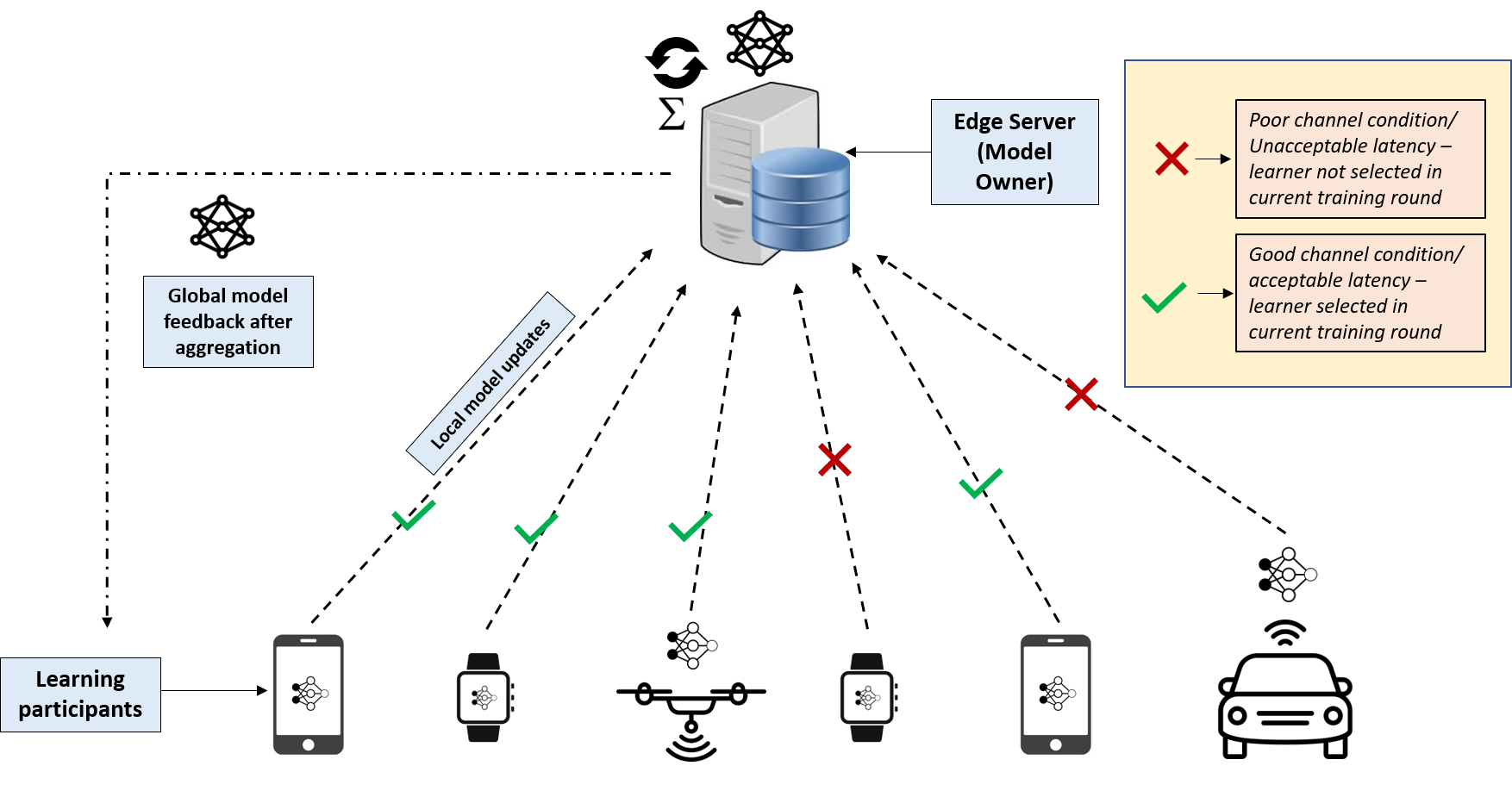}
    \caption{Federated Learning under wireless channel constraints}
    \label{FL_wireless_channel_constraints_fig}
\end{figure*}

In the following description, a \textit{local} model refers to the model trained locally at each participating end device, while the \textit{global} model refers to the model aggregated by the edge server. As shown in Figure~\ref{FL_wireless_channel_constraints_fig}, we consider an FL network with multiple end devices being served by a single edge server. Denote $I$ as the set of all end devices, the size of which is $N$. Initially, the edge server publishes the global model parameters to all the participating devices. In each training round, the participant devices use their respective local datasets and compute updates to their local models based on new sensed data, which are subsequently sent to the edge server. At the end of each training round, the edge server aggregates these local updates, derives and sends back the updated global model parameters. In practice, most often the aggregation simply amounts to a weighted average~\cite{McmahanFedML2017}, however this may vary. The local model update and parameter aggregation are repeated until the global loss function (typically an average of local loss functions) converges.

To train a global model repeatedly, end devices have to complete their local training tasks and transfer the local model parameters within each training around. We denote $t = T_{\text{upd}}$ as the duration of a training round, after which the edge server commences global model aggregation. If the model parameters of an end device cannot meet the constraint of $T_{\text{upd}}$, the device has no contributions on the global training task. In addition, realizing that training machine learning models is typically resource-intensive, it is imperative to strive for optimal operation of the training task to avoid resource wastage. These resources may be in the form of bandwidth, time, energy, cost, etc. In this context, let $L_{\text{max}}$ be the maximum volume of local model parameters that the edge server can process. Such a constraint is very relevant, since a practical requirement for deploying FL would be to minimize communication delay, analogous to minimizing control plane signaling latency in packet switched networks. Also, let \(\tau_{\text{comp},i}\) be the delay incurred in computing local model updates at participating device $i$ in a given training round. This delay is dictated by the on-device processing capabilities. For a particular device, given its computational intensity $\alpha$ at an end device (i.e., the number of CPU (Central Processing Unit) cycles needed to process one bit of data sensed by the device) and its computation capacity $C_i$ (in CPU cycles/second), the computing delay $\tau_{\text{comp},i}$ can be expressed as $\tau_{\text{comp},i} = \alpha D_i/C_i$. 
Similarly, let \(\tau_{\text{comm},i}\) be the communications delay incurred in transferring the local model update from device $i$ to the edge server.
Note that this communications delay is a function of the channel condition $h_i$ from the device to the edge server, the bandwidth $B_{\text{ch}}$ used by this connection, and the data size of its local model update $l_i$. Thus, $\tau_{\text{comm},i}$ can be defined as $\tau_{\text{comm},i} = l_i/r_i$, where $r_i$ is the data rate supported by this connection.
On the other hand, as end devices train their local models based on the local datasets, the size of a dataset directly impacts the accuracy of an end device's model parameters. To better reflect the contribution of an end device in deriving the global model, we define \(w_i = D_i/\sum\limits_{i=1}^{N} D_i\) as the weight of each device which is proportional to the size of its own local dataset. Denote $X_i$ as the indicator variable that represents the selection of device $i$ for the current training round. Therefore, the problem of optimizing the global model accuracy can be formulated as follows:

\begin{equation}
\begin{aligned}
     \mathop{\max} \limits_{{X_i}} &\sum_{i=1}^N {w_i}{X_i}\\
 s.t. \sum_{i=1}^{N} {X_i}{l_i} &\leq L_{\text{max}}, \ \forall~i\in I, \\
  {X_i} & = \{0, 1\},\  \forall~i\in I, \\
 \tau_{\text{comp},i} + \tau_{\text{comm},i} &\leq T_{\text{upd}},\ \forall~i\in I.
\end{aligned}
\label{Knapsack_eqn}
\end{equation}

Here, the first constraint imposes the total collected model parameters to be no more than the capacity of the edge server, in terms of the maximize size of parameters that can be processed. The second constraint indicates that each end device can be selected or removed by the edge server. The third constraint imposes the local training time and communications delay to be less than the training round time. Note that the size of the local model update $l_i$, can differ from one participant to another. This is especially true in cases where the participating devices are responsible for training a different part of the same global model. This assumption also aligns well with non-IID assumption of the local datasets. Figure~\ref{FL_wireless_channel_constraints_fig} depicts the participant selection process in one training round of FL. 

The above participant selection problem can be transformed to a well-known $0/1$ Knapsack problem, which poses challenges in achieving the optimal solution due to the high computational complexity especially for a large number of end devices. To arrive at an efficient solution, we design a greedy algorithm to solve the problem in polynomial-time such that the objective value $\sum_{i=1}^M {w_i}{X_i}$ is maximized, thereby reflecting the accuracy of the global model. The detailed procedure of the algorithm is described as follows. In each training round, we first obtain a set of devices that meet the delay constraint described in Equation~\eqref{Knapsack_eqn}. Subsequently, we calculate the weight $z_i = w_i/l_i$ for all devices in the set, and sort them in descending order of their weights. Subsequently, the end device with the highest $z_i$ is iteratively selected by the edge server until the edge server cannot accommodate more end devices due to the constraint on its capacity $L_{\text{max}}$. In this manner, the accuracy of the global model in the edge server is improved by selecting such devices that collect more sensed data to train their local models.

\subsection{Simulation Results}
To further verify the performance of the designed algorithm,
we set up a simulation and compare it with a baseline algorithm (i.e., Best-SINR algorithm). The basic idea of the Best-SINR algorithm is to select devices with the best channel conditions, akin to the widely-used user association mechanism in wireless networks. In this simulation, we consider three categories of devices, namely, smartphones, vehicles and low-power IoT sensors, where the proportion of each category of devices is pre-defined. To be specific, for any given $N$, we assign $50\%$ of the devices as smartphones, $30\%$ as vehicles and $20\%$ as IoT sensors. The devices are randomly distributed in a $1$ $\text{km}^2$ area. Their training dataset sizes are set according to a normal distribution (i.e., $\mathcal{N}$($\mu$, $\sigma^{2}$)), with each device category having a specific mean value $\mu_{D_i}$. Meanwhile, we choose $l_i$ for each device based on a normal distribution as well.
The transmission power $P_t$ of each device is set at $200$ mW, and we employ the following pathloss (PL) model from $3$GPP~\cite{pathloss_LTE}, i.e., $PL \text{(dB)} = 128.1 + 37.6\log_{10} d$, where $d$ is the distance in kilometers. The remaining simulation parameters have been defined in Table~\ref{simulation_parameters_table}.

\begin{table*}
\centering
\caption{Simulation Parameters - Definitions and Values used}
\begin{tabular}{llc} 
\toprule
\textbf{Parameter} & \textbf{Definition}                            & \textbf{Value}                                   \\ 
\hline
Service area       & Coverage area served by Edge Server            & $1$ km\textsuperscript{2}                                            \\
$N$                  & Number of users present in Service area        & $300$ (Fig.~\ref{results_fig1_Obj_vs_Lmax})                                             \\
$\alpha$              & Computational intensity                  & $0.5$ (CPU cycles/bit)                             \\
$T_{\text{upd}}$               & Global model update period                     & $120$ ms                                           \\
$L_{\text{max}}$               & Aggregator Capacity                           & $1$ MB (Fig.~\ref{results_fig2_Obj_vs_N})                                             \\      
$B_{\text{ch}}$                & Channel bandwidth per user                     & $180 \text{kHz}$ $\triangleq$ \text{1 RB (Resource Block)}  \\
$P_t$                 & Transmit Power per user                        & $200$ mW                                           \\
$N_0$                 & Noise Spectral Density                         & $-174$ dBm/Hz                                      \\
$\mu_{D_1}$             & Mean size of training dataset for smartphones  & $150$ kB                                           \\
$\mu_{D_2}$             & Mean size of training dataset for vehicles     & $250$ kB                                           \\
$\mu_{D_3}$             & Mean size of training dataset for IoT sensors  & $100$ kB                                           \\
$\sigma_D$              & Standard deviation of training dataset for all device types   & $20$ kB                              \\
$\mu_{l}$              & Mean parameter update length                   & $10$ kB                                            \\
$\sigma_{l}$           & Standard deviation of parameter update length     & $2$ kB                                           \\
${C_1}$             & Computational capacity of smartphones & \num{1e6} CPU cycles/sec                                              \\
${C_2}$             & Computational capacity of vehicles & \num{2e6} CPU cycles/sec                                             \\
${C_3}$             & Computational capacity of IoT sensors & \num{5e5} CPU cycles/sec                                           \\
\bottomrule
\end{tabular}
\label{simulation_parameters_table}
\end{table*}

 

\begin{figure}[!h]
    \centering
    \includegraphics[width=1\linewidth, height=0.75\linewidth]{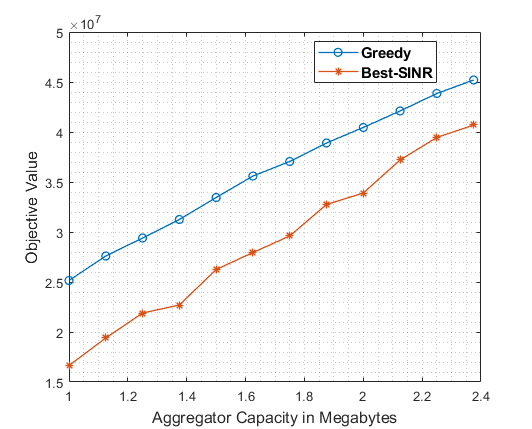}
    \caption{Objective Value vs Aggregator Capacity $L_{\text{max}}$}
    \label{results_fig1_Obj_vs_Lmax}
\end{figure}

Figure~\ref{results_fig1_Obj_vs_Lmax} illustrates the impact of the edge server's aggregator capacity $L_{\text{max}}$ on the objective value. It can be seen that the greedy algorithm is able to consistently achieve a higher objective value compared to the baseline over a wide range of aggregator capacity $L_{\text{max}}$. This is attributed to the fact that the greedy algorithm prefers to select devices with a larger local training dataset and thus ensures a higher accuracy of the global model for a given capacity limitation of the aggregator at the edge server. In contrast, the Best-SINR algorithm simply selects devices in its vicinity, and thus cannot guarantee the accuracy of the global model.

\begin{figure}[!h]
    \centering
    \includegraphics[width=1\linewidth, height=0.75\linewidth]{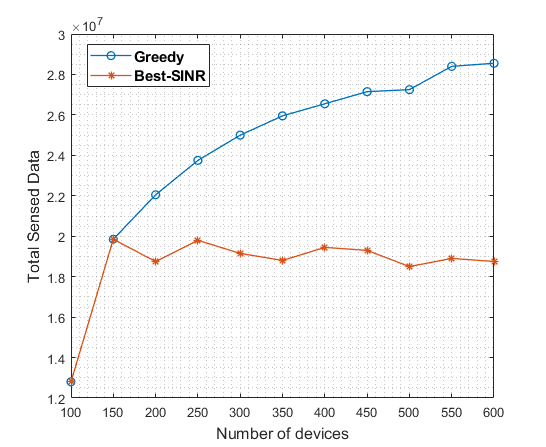}
    \caption{Total sensed data (in bytes) used for aggregation vs Number of devices $N$}
    \label{results_fig2_Obj_vs_N}
\end{figure}

Furthermore, in Figure~\ref{results_fig2_Obj_vs_N}, we analyze the impact of the number of participating end devices on the total sensed data that contributes towards the global model. It can be seen that the total sensed data of the two algorithms are similar when the number of participating devices is small. This is because in this case, all feasible devices satisfying the delay constraint are accommodated by the aggregator in the edge server. As the number of devices increases, the greedy algorithm is able to accommodate a much larger amount of data sensed by the participating devices and thus, has a much better global model accuracy while the performance of the Best-SINR algorithm reaches an approximate upper bound in terms of the amount of user data it can contribute to the global model. This phenomenon can be attributed to the fact that the greedy algorithm continues to track the devices with larger datasets while the Best-SINR approach selects a certain number of devices in its vicinity constrained by the aggregator capacity even when the total number of devices increases. Overall, the greedy algorithm takes into account each device's dataset size and results in a more accurate update to the global model in each training round, leading to more optimized FL. 

\section{Future Research Challenges}

\subsection{Optimizing Learning Delay} 
The training delay of the global model in FL will be a critical metric in real-time learning applications. In a given network, there may be certain devices that are constrained by their wireless channel conditions, while other devices may have energy constraints, thus limiting their computational capabilities. In this context, given a required accuracy of the global model, the selection policy of end devices accounting for their energy constraints will play an important role in minimizing the overall training delay. Therefore, the above problem of balancing between global model accuracy and increasing operational longevity of user devices remains an open challenge.

\subsection{Participation and Privacy Considerations} Ensuring maximum end device participation in the shared global model training remains a challenge when a majority of participants are supported on potentially unreliable wireless channels. Additionally, battery-powered IoT devices may also implement policies specific to distributed learning to conserve energy, thereby not participating in certain rounds of training for a given task.
Furthermore, the hallmark feature of FL is the guarantee of local user data privacy. However, adversaries may still be able to extract sensitive information from model updates alone~\cite{Giuseppe2015}. Although solutions based on differential privacy~\cite{AbadiDP2016} and secure aggregation~\cite{Bonawitzsecure2016} have been proposed, securing local model updates may come at the expense of model accuracy.
Therefore, learner selection strategies in general need to strike a balance between maximum user participation for higher training accuracy and respecting the security concerns of end devices.
    


\subsection{Convergence studies for Asynchronous FL} Asynchronous FL is an approach allowing participants to join the training phase while it is in progress, as opposed to Synchronous FL that is prone to the straggler effect~\cite{JalaliradAsyncFL2019}. Although Synchronous FL is the preferred approach owing to better convergence guarantees~\cite{BonawitzFLscale2019}, the asynchronous model is better suited to real-world FL training scenarios with end devices being channel and energy-constrained. Convergence bounds and generalization abilities of popular algorithms such as distributed gradient descent need to be studied theoretically to arrive at convergence guarantees for convex as well as non-convex loss functions.  


\section{Conclusion}
This article presents a brief introduction to Multi-Access Edge Computing (MEC), its motivation and how Federated Learning (FL) together with MEC can play an important role in efficiently realizing learning and inference tasks under extremely stringent latency constraints, that are will be one of the critical constaints in the Beyond-$5$G vision. The conceptual foundations behind MEC and FL are introduced, accompanied by a brief state-of-the-art survey. Under the larger umbrella of FL, two inter-related optimization problems focusing on minimizing communications latency in FL tasks and maximizing FL training accuracy under wireless channel constraints and edge server limitations are formulated, along with a discussion on simulation results obtained via the introduced greedy algorithm. We also present a relevant sampling of current challenges and future directions for edge computing-assisted FL research for Beyond-$5$G networks.

\bibliographystyle{IEEEtran}
\bibliography{IEEEabrv,ref}

\end{document}